\documentclass[twocolumn,showpacs,preprintnumbers,amsmath,amssymb]{revtex4}
\usepackage{graphicx}
\usepackage{dcolumn}
\usepackage{bm}
\usepackage[tight]{subfigure}

\begin{document}
\title{Kondo-transport spectroscopy of single molecule magnets}
\author{C. Romeike$^{(1)}$}
\author{M. R. Wegewijs$^{(1)}$}
\author{W. Hofstetter$^{(2)}$}
\author{H. Schoeller$^{(1)}$}
\affiliation{
(1) Institut f\"ur Theoretische Physik A, RWTH Aachen, 52056 Aachen,  Germany \\
(2) Institut f\"ur Theoretische Physik, J. W. Goethe-Universit\"at Frankfurt, 60438 Frankfurt am Main, Germany
}

\begin{abstract}
  We demonstrate that in a single molecule magnet (SMM) strongly
  coupled to electrodes the Kondo effect involves all magnetic
  excitations.
  This Kondo effect is induced by the quantum tunneling of the
  magnetic moment (QTM).
  Importantly, the Kondo temperature $T_K$ can be much larger than the
  magnetic splittings.
  We find a strong modulation of the Kondo effect as function of the
  transverse anisotropy parameter or a longitudinal magnetic field.
  For both integer and half-integer spin this can be used for an
  accurate transport spectroscopy of the  magnetic states
  in low magnetic fields on the order of the easy-axis anisotropy parameter.
  We set up a relationship between the Kondo effects for successive
  integer and half-integer spins.
\end{abstract}

 \pacs{
   72.10.Fk,  
   75.10.Jm,  
   75.30.Gw, 
   75.60.Jk   
}

\maketitle

\begin{figure}
  \includegraphics[scale=0.33]{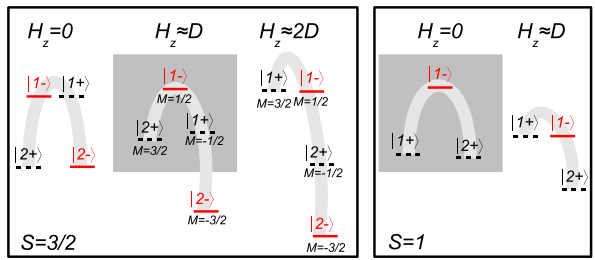}
  \caption{\label{fig:levels}
    Energy diagram for $H_\text{M}$ with $B_2 \ll D$ without (left) and
    with (right) magnetic field.
    The two subsets of eigenstates of different magnetic symmetry are indicated
    by full and dashed bars.
  }
\end{figure}

\begin{figure}
  \subfigure[]{\includegraphics[width=0.48\linewidth]
    {fig2a.eps}
    \label{fig:NRG_2S7_B2}}
  \subfigure[]{\includegraphics[width=0.48\linewidth]
    {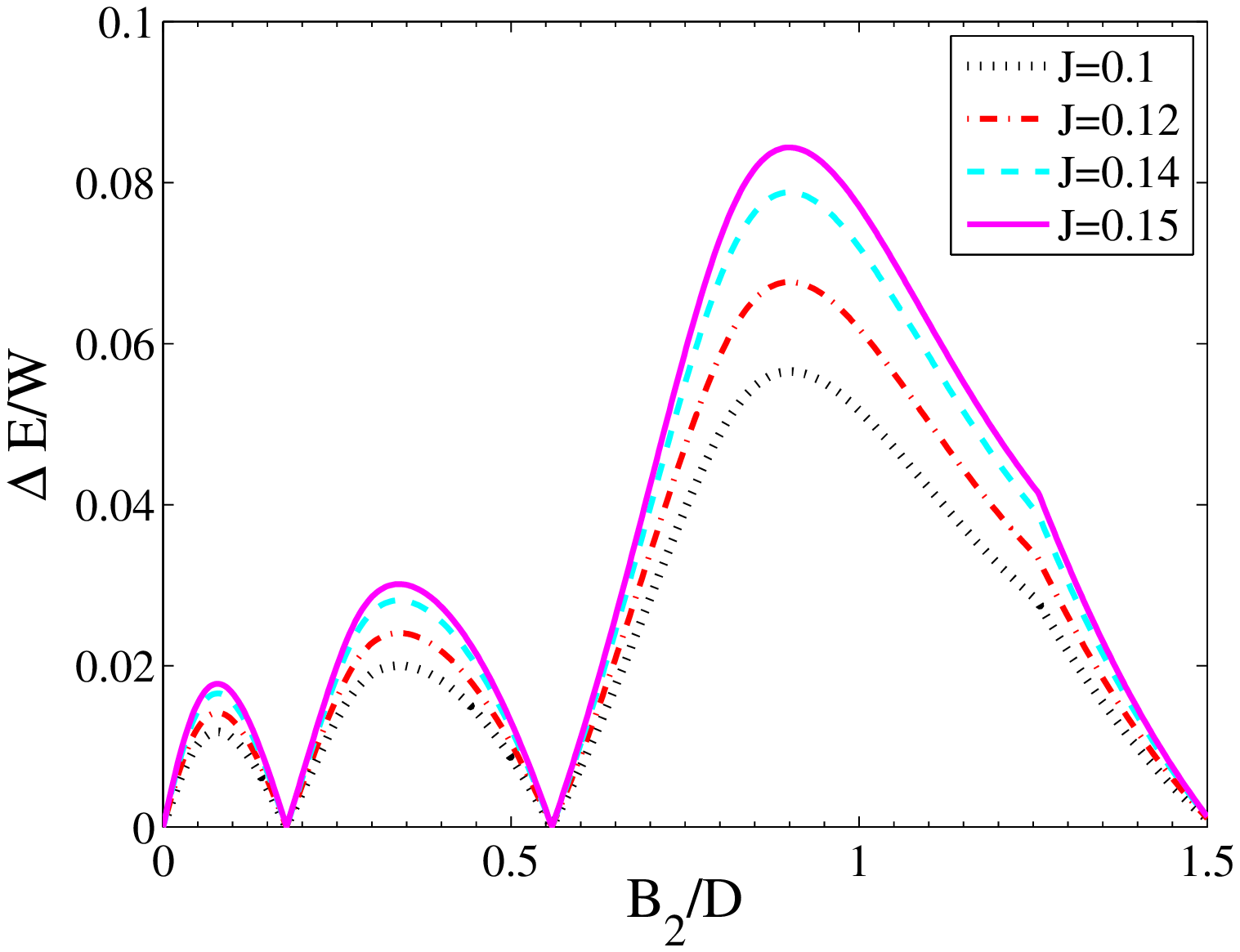}
    \label{fig:2site_2S7_B2}}
  \caption{
    Dependence of $T_K$ on the transverse anisotropy $B_2/D$:
    (a) $T_K$ from the full NRG calculation, 
    (b) spin binding energy estimate $\Delta E$.
    Parameters: $S=7/2$, $D=5*10^{-5}W$.
  }
\end{figure}

\begin{figure}
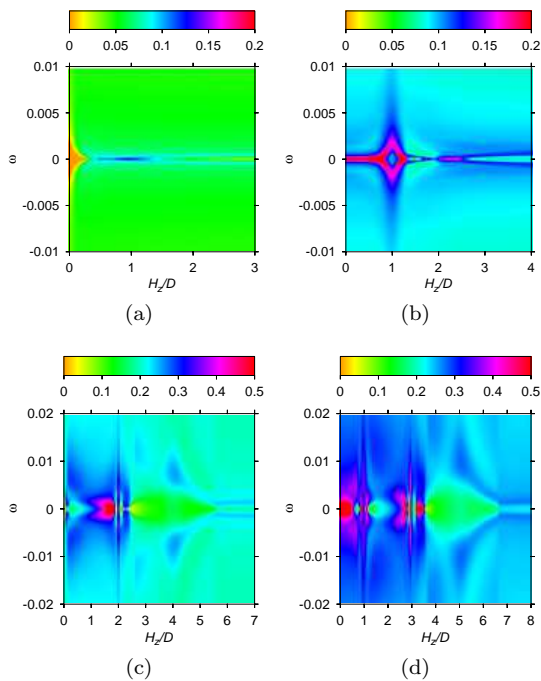

  \subfigure[]{\includegraphics[width=0.4\linewidth]{fig3a.ps}
    \label{fig:Hz_2S2}}
  \subfigure[]{\includegraphics[width=0.4\linewidth]{fig3b.ps}
    \label{fig:Hz_2S3}}
  \subfigure[]{\includegraphics[width=0.4\linewidth]{fig3c.ps}
    \label{fig:Hz_2S6}}
  \subfigure[]{\includegraphics[width=0.4\linewidth]{fig3d.ps}
    \label{fig:Hz_2S7}}
  \caption{(Color online)
    Spectral function, qualitatively equivalent to $dI/dV_\text{bias}$ as function of $H_z/D$
    and frequency $\omega$ (a) for smallest non-trivial integer spin $S=1$,
    (b) half-integer spin $S=3/2$, and 
    (c) large integer $S=3$, (d) half-integer spin $S=7/2$.
    Parameters: $J=0.15W, D=5*10^{-5} W$ and $B_2/D=0.1$.
  }
\end{figure}

{\em Introduction.---}
Single molecule magnets (SMMs) allow the study of quantum phenomena on a mesoscopic
scale, namely the quantum tunneling of the magnetic moment.
SMMs such as Mn$_{12}$ or Fe$_{8}$ have attracted intense
experimental and theoretical investigation~\cite{gattorev03} in recent years.
Due to weak intermolecular interaction molecular crystal properties
can be assigned to single molecules described by a large spin
($S>1/2$), and easy-axis and transverse anisotropies.
Recently, two groups~\cite{Heersche,Jo06} have trapped
a \emph{single} molecule magnet (Mn$_{12}$) in a three terminal transport
setup and measured transport through the single molecule.
Electron transport fingerprints due to
both sequential~\cite{Heersche,Jo06} and inelastic co-tunneling~\cite{Jo06}
processes were observed and associated with the molecular magnetic
states.
Theoretical works~\cite{Kim04,Romeike06a,Romeike06b} predicted
that fingerprints of magnetic quantum tunneling (QTM) can in
principle be identified in transport measurements in the charge
(sequential tunneling) as well as in the spin fluctuation (Kondo)
regime. Especially the latter regime of strong coupling to the electrodes
is of importance since the Kondo effect shows up as a 
sharp zero-bias anomaly with width given by the Kondo temperature
$T_K$.
The Kondo effect has been observed experimentally in many other
systems with small spins (e.g.  quantum
dots~\cite{Goldhaber98,Cronenwett98,Simmel99,Schmid00,vdWiel00} and
single molecules~\cite{Park02,Liang02}). 
For SMMs with \emph{half-integer spin} $S$ in \emph{zero magnetic field} 
it was shown~\cite{Romeike06a} that the Kondo effect
arises from a cooperation of both spin exchange processes with the
reservoirs and the intrinsic tunneling of the magnetic moment.
The strong coupling fixed point is of pseudo-spin-1/2 type.
In this limit the exchange coupling $J$ is weak, such that the
resulting Kondo temperature $T_{K}(J,\Delta) \ll \Delta$, where
$\Delta\lesssim 1$~meV is the scale of the anisotropy splittings
between the magnetic states.
 In this case, the Kondo effect involves only the two magnetic
\emph{ground states} of the SMM and does not occur for integer spin.
\\
In this Letter we study the Kondo effect in the experimentally more
favorable regime of strong exchange ($J \sim 0.1$ eV) where the Kondo
temperature becomes larger than the magnetic splittings
$T_K(J,\Delta) \gg \Delta$.
We find that \emph{excited} magnetic levels on the SMM, belonging to
different topological sectors (with respect to rotations around
the $z$-axis), become essentially involved in the Kondo effect.
In zero magnetic field and for half-integer spin the Kondo effect
can be modulated by changing the transverse anisotropy, resulting
in a sequence of Kondo effects associated with different magnetic
excited states.
A longitudinal magnetic field can induce two important new effects:
(1) For both integer and half-integer spin the Kondo effect is suppressed
at each anticrossing of magnetic levels belonging to the
\emph{same} topological sector.
The corresponding {sharp} magnetic field scale is determined by the
transverse magnetic anisotropy. Therefore the study of the
magnetic field dependence allows for an accurate spectroscopy of the
magnetic states.
This modulation of the Kondo effect allows for an experimental proof of
the existence of QTM in a \emph{single} molecule  \emph{in a transport junction} and
to determine the important microscopic parameters characterizing the
SMM in that setup.
The \emph{magnetic field induced} Kondo effect for integer spin SMMs
allows many molecular magnets like e.g. Fe$_8$ to be studied without
charging the molecule to obtain a half-integer spin.
(2) We find that the Kondo effects for successive integer and
half-integer spin $S$ display a close correspondence when shifted in
magnetic field energy by the easy-axis anisotropy parameter $D$.
\\
{\em Model.---}
We consider a SMM (Fig.~\ref{fig:levels}) in a transport setup where
the applied voltages, charging effects and low temperature suppress
single-electron processes~\cite{Romeike06a}. In the presence of a
longitudinal magnetic field the Hamiltonian reads
$H = H_\text{SMM} + H_\text{ex}+H_\text{res}$:
\begin{eqnarray}
  \label{eq:ham_mol}
  H_\text{SMM} &=& -D S_{z}^{2}
  + {\textstyle \frac{1}{2}} B_{2} \left (S_+^{2}+S_-^{2} \right)
  + H_z S_z \\
  \label{eq:ham_ex}
  H_\text{ex} &=&  J  \bm{S} \cdot \bm{s} \\
  \label{eq:ham_res}
  H_\text{res} &=&  \sum_{k \sigma} \epsilon_{k \sigma} a_{k \sigma}^{\dag} a_{k \sigma}
 \end{eqnarray}
where $S_{z}$ is the projection of the SMM's spin on the easy
$z$-axis, and $S_\pm = S_x \pm i S_y$.
The terms in Eq.~(\ref{eq:ham_mol}) describe, respectively,
the easy-axis magnetic anisotropy of the molecule, the transverse
anisotropy perturbation and the coupling to a magnetic field ($H_z$) along the
easy axis. The $g$-factors are absorbed into the magnetic field.
For simplicity we have taken a 2-fold rotation-symmetry axis, which is
dominant in many molecular magnets.
The Hilbert space of $H_\text{SMM}$ is split into two disjoint
topological subspaces $\sigma=\pm$ since the 2-fold rotational
symmetry about the $z$-axis is preserved under the longitudinal
magnetic field.
While for half-integer $S$ these spaces are spanned
by an equal number of $n_{+}=n_{-}=S+1/2$
basis-states $\lbrace | S,M \rangle \rbrace_{M=\sigma S,\sigma
(S-2),\ldots,-\sigma (S-3),-\sigma (S-1)}$,
for integer spin they are spanned
by $n_{+}=S+1$ states
$\lbrace | S,M \rangle \rbrace_{M=-S  ,-S+2,\ldots,S-2,S}$
for $\sigma=+$ and
by $n_{-}=S$ states
$\lbrace | S,M \rangle \rbrace_{M=-S+1,-S+3,\ldots,S-3,S-1}$
for $\sigma=-$.
The eigenstates $|l \sigma \rangle$ of $H_\text{SMM}$ with energies
$E_{l\sigma}$ are labeled by $l=1,2,..,n_{\sigma}$ in order of
decreasing energy in each subspace (see Fig.~\ref{fig:levels}).
Eq.~(\ref{eq:ham_ex}) describes the exchange coupling of the
molecular spin to the effective reservoir Eq.~(\ref{eq:ham_res})
(bandwidth $2W$ and constant density of states $\rho$).
The electronic states are labeled by $k \sigma$ and denote the
even combination of left and right physical electrode
states~\cite{Glazman88,Ng88}. The local electron spin in the reservoir is
$\bm{s} = \sum_{k k'} \sum_{\sigma \sigma'}
  a_{k\sigma}^{\dag} \bm{\tau}_{\sigma \sigma'}
  a_{k'\sigma'}/2$
where $\bm{\tau}$ is the Pauli-matrix vector.
The exchange coupling $J$
is induced by virtual electron tunneling processes and is
antiferromagnetic due to the strong energy- and charge quantization effects.
The Kondo spin-scattering of conduction electrons off the SMM
transfers charge between physical left and right electrodes and
can resonantly enhance the linear conductance.
The half-width at half-maximum of the resulting zero-bias differential
conductance peak at $T=0$ is the Kondo temperature $T_K$.
\\
For a qualitative discussion it is useful to change to the exact
representation of Eq.~(\ref{eq:ham_ex}) in the eigenbasis of the
molecular states $|l\sigma\rangle$:
\begin{equation}
  \label{eq:ham_proj}
  H_\text{ex} = \sum_{l l'}
    \left(
      \sum_{i=x,y} J^{i}_{l l'} P^i_{l l'} s_i
    +\,\sum_{\sigma}  J^z_{l l'\sigma}
    \,| l \sigma\rangle \langle l'\sigma | \,s_z
    \right)
    .
\end{equation}
The Hamiltonian Eq.~(\ref{eq:ham_proj}) describes the exchange
processes involving reservoir electrons and internal magnetic degrees
of freedom of the SMM.
This projection makes explicit that the QTM parameter
$B_2$ and the external magnetic field $H_z$
modulate the \emph{effective} couplings
through the matrix elements of the molecular spin operator $S$.
The first, transverse, term describes the spin-scattering involving a
pair of states $|l+\rangle$ and $|l'-\rangle$ in terms of
pseudo-spin-1/2 operators
$P^{x}_{l l'}, i P^{y}_{l l'}
=  (| l +  \rangle \langle l' - |
              \pm   | l' - \rangle \langle l + | )/2$
with effective exchange couplings
\begin{equation}
  \label{eq:jxy}
  J^{x/y}_{l l'}=J\,\langle l +  | S_{+} \pm S_{-} | l' - \rangle
  .
\end{equation}
In total, there are $n_{+}n_{-}$ such pairs of states from opposite
topological sectors.
The longitudinal couplings read
$J^z_{l l'\sigma}=J  \langle l \sigma   | S_{z} | l' \sigma \rangle$.
Longitudinal spin operators
$P^{z}_{l l'}= (| l +   \rangle \langle l  + |
               -| l' -  \rangle \langle l' - |)/2$
can only be introduced in a \emph{unique} way
if an approximative projection onto a single pair is made.
For instance, one recovers the projection of Ref.~\cite{Romeike06a} onto the
ground state pair for zero-field and half-integer $S$ by using
time-reversal symmetry and by truncating the excited states.
However, such a truncation is not valid in the regime of interest here:
the strong exchange coupling $J$ gives rise to a Kondo temperature
that can be larger than the magnetic splittings ($T_K \gtrsim 2SD$ for
$B_2 \ll D$). The Kondo effect can occur irrespective of whether the
pair of states $|l+\rangle$ and $|l'-\rangle$ are ground-
or excited states of the SMM and whether they are exactly degenerate
or not.
The Kondo effect therefore involves contributions from multiple
topological pairs.
This we confirmed explicitly by calculating the projection of the full
many-body ground state onto each molecular eigenstate using the NRG.
Thus the full expression for $H_{\text{ex}}$, Eq.(~\ref{eq:ham_proj}),
must be retained for which a \emph{unique} decomposition into a sum of
\emph{independent} pairs is not possible.
This is due to the fact that a single eigenstate of the SMM
$|l \sigma\rangle$ is paired with $n_{-\sigma} > 2$
states $|l'-\sigma\rangle$.
\\
{\em Physical picture.---}
Variation of either the QTM parameter $B_2$ or the magnetic field
$H_z$ results in anticrossings of magnetic levels.
As shown in Fig.~\ref{fig:levels} the variation of the magnetic field
leads to an alternating sequence of degeneracies between pairs of levels
which are in different subspaces (crossings) or in the same
subspace (anticrossing).
Importantly, this occurs in the low magnetic field energy window where
the Zeeman splittings are still smaller than $T_K$
($E_{l\sigma}-E_{l'\sigma'} < T_K$).
The effective couplings in Eq.~(\ref{eq:ham_proj}) are modulated
strongly at each anticrossing leading to a suppression of the Kondo
effect on a scale $\sim B_2$, as we now explain.
At an anticrossing states $|l \sigma \rangle$ and $| l'\sigma\rangle$
from the \emph{same} topological sector are close in energy and
strongly hybridize due to the transverse anisotropy.
Following two levels \emph{adiabatically} during an
anticrossing they interchange their role and one basis state picks up
a relative phase $\pi$.
For example, in Fig.~\ref{fig:levels}  the level $|1+\rangle$ is moving
upwards and the level $|2+\rangle$ downwards in energy  after the
anticrossing.
For \emph{each} topological pair involving one of the anticrossing
levels this leads to a sign change of one of the transverse couplings,
as can be seen easily from Eq.~(\ref{eq:jxy}).
The latter therefore vanish at the anticrossing and
the Kondo effect is expected to be suppressed.
This happens on a magnetic field scale proportional to the
tunnel splitting, i.e. will occur
as a sharp feature in the linear conductance as function of a
longitudinal magnetic field.
In a similar way, in zero field the variation of the QTM parameter
$B_2$ itself leads to a series of anticrossings, which for a
half-integer spin preserves the 2-fold Kramers degeneracy of all
levels.
This happens due to the non-uniform level spacing and the coupling of
states occurs in different orders of perturbation theory in $B_2$
(higher-lying levels hybridize more strongly for weak $B_2$).
\\
{\em Method.---}
We use Wilsons's NRG~\cite{Wilson75,Hofstetter00} to treat
non-perturbatively the full model $H = H_\text{SMM}+ H_\text{ex} +
H_\text{res}$.
As NRG input parameters we use number of states $N=1000$,
discretization $\Lambda=2$ and $D= 5 \times 10^{-5} W$.
We analyzed the NRG level flow as function of iteration
number $N_\text{iter}$ in order to determine the low-temperature fixed
point for $H_z=0$.
The Kondo temperature is defined as the energy scale
where the crossover to strong coupling takes place.
We also calculated the spectral function within
the $T$-matrix approach~\cite{Costi00}:
$
  A_\sigma (\omega) =
  -\frac{1}{\pi} \text{Im} T_\sigma (\omega+i \delta)
$
where the $T$-matrix is
$
  T_\sigma (\omega) =
  \langle \langle O_{\sigma}; O^\dag_{\sigma}  \rangle \rangle
$
with
$
 O_\sigma =
 \frac{J}{2} \left(
   c_{0,-\sigma} S^{-\sigma}+\sigma c_{0,\sigma} S_z
\right)
$
and $c_{0,\sigma}=\sum_k c_{k,\sigma}$ is the electron operator on the
first site of the Wilson chain.
The low temperature conductance
$G  =
 -(e^2/2) \sum_{\sigma} \int d \omega  A_\sigma(\omega) df(\omega)/d
 \omega$
is proportional to the spectral function, where $f(\omega)=1/(e^{\omega/T}+1)$.
The effects considered here are difficult to capture in a poor-man's
scaling approach due to the strong coupling and the many excited
states involved.
However, another useful guide to the full NRG
results in this regime is the \emph{spin binding
energy} $\Delta E$, obtained by diagonalizing exactly
the molecular Hamiltonian Eq.~(\ref{eq:ham_mol}) coupled by exchange
to a \emph{single} conduction electron spin.
This is similar to the first step of a NRG calculation
and will be denoted as the zero-bandwidth model.
The ground- to excited-state gap $\Delta E$ thus obtained
follows the modulation of $T_K$ by $B_2$ and $H_z$ accurately,
although the {scales} of $\Delta E$ and $T_K$ strongly differ.
Apparently, for a sizable range of the relevant exchange
strengths $J<W$ the zero-bandwidth estimate $\Delta E$ is
renormalized \emph{uniformly} by the coupling to the remaining
conduction band electrons.
This estimate breaks down, when the modulation involves a suppression
of $T_K$ far below the scale of the magnetic splittings: then the
renormalization becomes non-uniform and the variation may deviate from
the NRG.
\\
{\em Kondo effect due to excited states tuned by QTM.---}
We first consider the case of zero magnetic field.
For integer spin this corresponds to an anticrossing: the eigenstates
are paired to nearly degenerate states from the same subspace which
are splitted by the transverse anisotropy.
Consequently, as discussed above the Kondo effect is suppressed,
resulting even in a \emph{dip} in the spectral function, see below.
In contrast, for half-integer spin many topological pairs are
degenerate (crossing) due to time-reversal symmetry, see
Fig.~\ref{fig:levels}.
The NRG converges to the spin-1/2 strong coupling fixed point,
indicating a complete screening of the magnetic degrees of freedom.
As shown in Fig.~\ref{fig:NRG_2S7_B2} the Kondo temperature shows an
oscillatory dependence on $B_2/D$. The number of
oscillations for $0 \leq B_2 \leq D$ is $S-1/2$.
When decreasing the exchange coupling $J$ the smallest peaks disappear
first, leaving only the monotonic increase of $T_K$ up to the broad
maximum centered at $B_2 \sim D$ in the limit where the magnetic
excitations can be neglected~\cite{Romeike06a}.
The spin binding energy $\Delta E$ captures this dependence on
$B_2/D$, as shown in Fig.~\ref{fig:2site_2S7_B2}.  Interestingly, we
find that the nature of the ground state of the zero-bandwidth model
changes with each oscillation which can in turn be related to an
anticrossing of magnetic states on the SMM.
For instance, for weak $B_2 \ll D$ near the lowest peak in
Fig.~\ref{fig:2site_2S7_B2} the highest \emph{excited}
 Kramers doublets are strongly mixed into the zero-bandwidth model ground state.
In contrast, for $B_2 \sim D$ near the highest peak in
Fig.~\ref{fig:2site_2S7_B2} the ground doublet dominates
in the ground state.
This demonstrates that the observed strong coupling Kondo fixed point in the
NRG indeed originates from a screening of magnetic degrees of freedom
involving excited magnetic states of the SMM.
Which of these excited states are important depends on $B_2/D$.
\\
{\em Kondo-spectroscopy of SMMs.---}
We now focus on the most relevant case of fixed weak QTM ($B_2\ll D$)
and vary the longitudinal magnetic field $H_z$.
For many magnetic field values the spectral density shown in
Figs.~\ref{fig:Hz_2S2}-\ref{fig:Hz_2S7} displays a zero-bias Kondo
resonance. The corresponding many-body ground state found in the NRG
at the strong coupling fixed point is non-degenerate.
This peak is strongly modulated at avoided crossings of states from
the same subspace at $H_z/D \approx 2k$ ($2k+1$), $k=1,2,\ldots$ for
integer (half-integer) spin.
As discussed above, this is related to the strong suppression of
effective exchange couplings on the magnetic
field scale $B_2$ near an anticrossing.
In contrast, the spectral function varies smoothly at level
crossings located at $H_z/D \approx 2k+1$ ($2k$) for integer
(half-integer).  In this way the two scales $D$ and $B_2$ can
be identified in the magnetic field dependence of the linear
conductance.
For integer spin $S=1$ the spectral function in Fig.~\ref{fig:Hz_2S2}
shows a dip for $H_z=0$ (anticrossing of $|1+\rangle$ and
$|2+\rangle$).
At fields $H_z \sim B_2$ a Kondo peak is \emph{induced} which reaches
maximal height at $H_z = D$ (crossing of $|1+\rangle$ and
$|1-\rangle$).
This Kondo peak is subsequently suppressed and splitted due to
the Zeeman level shifts, see Fig.~\ref{fig:levels}.
For half-integer spin $S=3/2$ the zero-frequency peak at $H_z=0$ in
Fig.~\ref{fig:Hz_2S3} is broadened with increasing $H_z$.
This broadening is maximal at $H_z=D$ and is also captured by the
$H_z/D$ dependence of the spin-binding energy $\Delta E$ (not shown).
The full NRG result in addition shows a sharp dip superimposed on the
broad peak at $H_z = D$ (anticrossing of $|1+\rangle$ and $|2+\rangle$).
At this point the situation is very similar to the $S=1$ case,
provided one neglects the contribution of the low lying state
$|2-\rangle$ and reduces the magnetic field by $D$,
see Fig.~\ref{fig:levels}.
Indeed, increasing the field further to $H_z=2D$, the Kondo effect
reappears precisely as for $S=1$ near $H_z=D$.
For larger spin the physics is qualitatively the same: the spectral
function in Figs.~\ref{fig:Hz_2S6},~\ref{fig:Hz_2S7} displays
\emph{several} dips of the Kondo peak close to each anti-crossing.
The correspondence pointed out above is more general: the linear
conductance of a SMM with spin $S$ at magnetic field $H_z$ corresponds
qualitatively to that of a SMM with reduced spin $S-1/2$ and reduced
magnetic field $H_z-D$.
This can be shown using the Hamiltonian Eq.~(\ref{eq:ham_proj}) in the
molecular eigenbasis.
For fixed $J$ the correspondence in the conductance
is most obvious for subsequent $S$ values.
The conductance for spin $S$ near $H_z=nD$ ($n=1,2,\ldots,2S-1$) can
even be compared with that for $S-n/2$ around \emph{zero-field} by
iterating $n$ times, which we have checked.
However, a clear correspondence appears only
when $J$ is properly adjusted.
\\
{\em Conclusion.---}
Magnetic field Kondo transport spectroscopy can determine the absolute
value of the spin and the magnetic parameters of a SMM \emph{in a
transport junction}.
The strong Kondo effect in SMMs discussed in this work can be accessed
experimentally by increasing the exchange coupling.
In STM setups one can reduce the distance to the
molecule, or change the molecular geometry by a bias-voltage
pulse~\cite{Iancu06}.
In 3-terminal measurements~\cite{Heersche,Jo06} one can tune the
exchange coupling with the gate voltage: $J \propto 1/|V^{*}_g-V_g|$.
In addition, both the charge- and spin- state can be changed
by tuning the gate voltage to opposite sides of the charge degeneracy
point  $V_g=V^{*}_g$.
If a single-electron transport current~\cite{Romeike06b} is observed
at this point c.f.~\cite{Heersche,Jo06}, $S$ only changes by 1/2.
Then the predicted clear correspondence between the linear conductance
for subsequent integer and half-integer spin values
provides an additional check on the physics.
Importantly, the spectroscopy can be done at temperatures above the
energy scale of the magnetic splittings and requires only
low magnetic fields which do not destroy the Kondo effect by Zeeman
splittings.
Finally, we have checked that small fixed transverse magnetic field
perturbation does not destroy the reentrant behavior of the Kondo
effect as the longitudinal field is varied. 
\\
We acknowledge P. Nozi\`eres, H. Park and H. van der Zant for discussions
and financial support through the EU RTN Spintronics program
HPRN-CT-2002-00302 and the FZ J\"ulich via the virtual institute IFMIT.
\bibliographystyle{apsrev}

\end{document}